%% file: CTT_IT.tex
\documentclass[10pt,twocolumn]{IEEEtran}
\usepackage{amsmath,amssymb,amsthm, times, epsfig}
\usepackage[T1]{fontenc}

\usepackage[mathcal]{euscript} 

\newcommand{\asynclev}{A}

\newcommand{\cA}{\mathcal{A}}
\newcommand\cX{\EuScript{X}}
\newcommand\cY{\EuScript{Y}}

\newcommand{\cC}{\EuScript{C}}
\newcommand{\cD}{d}
\newcommand{\C}{\bm{C}}
\newcommand{\cE}{\EuScript{E}}

\newcommand{\cP}{\EuScript{P}}

\newcommand{\cT}{\EuScript{T}}

\newcommand{\ck}{\text{\it k}}
\newcommand{\cK}{\text{\it K}}

\newcommand{\bm}[1]{\mbox{\boldmath{$#1$}}}

\newcommand{\R}{\bm{R}}

\newcommand{\pr}{{\mathbb{P}}}
\newcommand{\ex}{{\mathbb{E}}}
\newcommand{\E}{{\EuScript{E}}}
\newcommand{\openone}{\leavevmode\hbox{\small1\normalsize\kern-.33em1}}

\newtheorem{thm}{Theorem}

\newtheorem*{cor}{Corollary}
\newtheorem*{rem}{Remark}

\newtheorem{defs}{Definition}
\newtheorem{fact}{Fact}
\DeclareMathOperator{\poly}{poly}

\title{Asynchronous Capacity per Unit Cost}
\author{Venkat Chandar,  Aslan Tchamkerten, and David Tse
\thanks{This work was supported in part by an Excellence Chair Grant
from the French National Research Agency (ACE project).   This work
was presented in part at the IEEE International Symposium on Information
Theory, Austin (Tx), USA, June 2010.} \thanks{V.~Chandar
is with MIT Lincoln Laboratory, Lexington, MA 02420, USA. Email:
vchandar@mit.edu.} \thanks{A.~Tchamkerten is with the
Department of Communications and Electronics, Telecom
ParisTech, 75634 Paris Cedex 13, France.
Email: aslan.tchamkerten@telecom-paristech.fr.}    \thanks{D.~Tse is with the
Department of
Electrical Engineering and Computer Sciences, University of California at
Berkeley, Berkeley CA 94729-1770, USA.
Email: dtse@eecs.berkeley.edu.}}

\begin{document}

\maketitle

{\begin{abstract} The capacity per unit cost, or, equivalently, the minimum cost
to
transmit one bit, is a well-studied quantity under the assumption of full
synchrony between the transmitter and the receiver. In many applications, such
as sensor networks, transmissions are very bursty, with amounts of bits
arriving infrequently at random times. In such scenarios, the cost of acquiring
synchronization is significant and one is interested in the fundamental limits
on communication without assuming {\em a priori} synchronization. In this
paper, the minimum cost to transmit $B$ bits of information asynchronously is
shown to be equal to $(B + \bar{H}) \ck_{\rm sync}$, where $\ck_{\rm sync}$ is
the synchronous minimum cost per bit, and where $\bar{H}$ is a measure of
timing uncertainty equal to the entropy for most reasonable arrival time
distributions. This result holds when the transmitter can stay idle at no
cost and is a particular case of a general result which holds for arbitrary
cost functions.
\end{abstract}}


\normalsize
\section{Introduction}
\label{intro} Synchronization is an important component of any
communication system. To understand the cost of synchronization,
it is helpful to divide applications into
two rough types. In the first type, transmission of data happens on
a continuous basis. Examples are voice and video. The cost of
initially acquiring synchronization, say by sending a pilot
sequence, is relatively small in such applications because the cost
is amortized over the many symbols transmitted.  In the second
type, transmissions are very bursty, with  amounts of data
transmitted once in a long while. Examples are sensor networks with
sensor nodes transmitting measured data once in a while. The cost of
acquiring synchronization is relatively more significant in such
applications because the number of bits transmitted per burst is
relatively small.

What is the fundamental limitation due to the lack of {\em a
priori} synchrony between the transmitter and the receiver in
bursty communication? While there has been a lot of research on
specific synchronization algorithms, this question has only
recently been pursued \cite{CTW,TCW,TCWj2}. In their model,
transmission of a message starts at a random time unknown to the
receiver. The performance measure is the data rate: the number of
bits in the message divided by the elapsed time between the
instant information starts being sent and the instant it is
decoded.

The data rate is a sensible performance metric for bursty
communication if the information to be communicated is
delay-sensitive. Then, maximizing the data rate is equivalent to
minimizing the time to transmit the burst of data. In certain
applications, however, the allowable delay may not be so tightly constrained,
so the data rate is less relevant a measure than the {\em energy}
needed to transmit the information. In this case, the minimum energy
needed to transmit one bit of information is an appropriate
fundamental measure. Thus, we are led to ask the following question:
what is the impact of asynchrony on the minimum energy needed to
transmit one bit of information?

This type of question falls into the general framework of {\em
capacity per unit cost} \cite{G4,Ve2}, where one is interested in
characterizing the maximum number of bits that can be reliably
communicated per unit cost of using the channel. Consider the
following modification of the formulation in \cite{TCW,TCWj2} to
study asynchronous capacity per unit cost.

There are $B$ bits of information which needs to be communicated.
The number $B$ can be viewed as the size of a burst in the above
scenario, with consecutive bursts occurring so infrequently that we
can consider each burst in complete isolation. The $B$ bits are
coded and transmitted over a memoryless channel using a sequence of
symbols that have costs associated with them. The rate $\bm{R}$ per
unit cost is the total number of bits divided by the cost of the
transmitted sequence.

The data burst arrives at a {\emph{random symbol time}} $\nu$, not
known {\em a priori} to the receiver. Without knowing $\nu$, the
goal of the receiver is to reliably decode the information bits by
observing the outputs of the channel. Although the receiver does not
know $\nu$, we assume that both the transmitter and the receiver
know that $\nu$ lies in the range from $1$ to $A$. The integer $A$
characterizes the asynchronism level or the timing uncertainty
between the transmitter and the receiver. At all times before and
after the actual transmission, the receiver observes pure noise.
The noise distribution corresponds to a special ``idle symbol'' $\star$
being sent across the channel.

The main result in this paper is a single-letter characterization of
the asynchronous capacity per unit cost, or, equivalently, the
minimum cost to transmit one bit of information. Under the further assumption
that the
idle symbol $\star$ is allowed to be used in the codewords and has zero cost,
the
result simplifies and admits a very simple interpretation: the minimum cost to
transmit $B$ bits of information
asynchronously is
\begin{equation}
\label{eq:min_cost}
(B + \log A) \ck_{\rm sync},
\end{equation}
where $\ck_{\rm sync}$ is the minimum cost to transmit one bit of
information in the synchronous setting.\footnote{In this paper, all logarithms
are taken to base $2$.} Thus, the timing uncertainty
imposes an additional cost of  $\ck_{\rm sync} \log A$ as
compared to the synchronous setting. Note that this result implies
that the additional cost is significant only when
the parameter $\log A$ is at least comparable to $B$.

Even though we do not have a {\em stringent} requirement on the
delay from the time of data arrival to the time of decoding, a
meaningful result cannot be obtained if there is {\em no} constraint
at all. This can be seen by noting that the transmitter could always
wait until the end of the arrival time interval (at time $A$) to
transmit information. Then, there would no price to pay for the
timing uncertainty since communication would {\it{de facto}} be synchronous.
However, the delay incurred would be very large if
$A$ is very large. To avoid this undesirable situation, we impose the
constraint that the delay should be
{\em linear} in $B$. A delay linear in $B$ is a natural constraint
since it is of the same order as the delay incurred in the
synchronous setting \cite{Ve2}. The expression \eqref{eq:min_cost}
is the minimum cost achievable by any scheme subject to this delay
constraint. Given this constraint, the start time of information
transmission is highly random to the receiver and the additional
cost is the cost needed to construct codewords that allow a decoder
to resolve this uncertainty.

What happens when longer delays are allowed? First, we show that
performance cannot be improved beyond \eqref{eq:min_cost} within the
broad class of coding schemes whose delays are  {\em
sub-exponential} in $B$. Second, we show that when the allowable
delay $d$ scales exponentially with $B$ (but is no larger than $A$, for
otherwise
the situation reduces to the synchronous setting mentioned above),
the minimum cost to transmit $B$ bits can be further reduced to
$$ \left(B + \log \frac{A}{d}\right)
\ck_{\rm sync}.$$ Thus, in this more general case,
the impact of asynchronism is significant when $\log (A/d)$
is at least of the order of $B$.

The above results are all proved under a uniform distribution on the arrival
time $\nu$. They can be generalized to a broad class  of other distributions,
with $\log A$ replaced by a quantity $\bar{H}$, which equals the entropy for
most reasonable distributions.

It is worth mentioning that the asynchronism studied in this paper
is due entirely to the random arrival time of the data and the
desire to deliver that data within a certain delay constraint. One
can think of this as {\em source} asynchronism. There is another
type of asynchronism due to the lack of a common clock between the
transmitter and the receiver. One can think of this as an example of
{\em channel} asynchronism. We do not consider this type of
asynchronism here. Hence, throughout the paper, we will assume both
the transmitter and the receiver have access to a common clock. An
interesting future direction would be to study the combined effect
of source and channel asynchronism.

\section{Model and Performance Criterion}\label{moper}
Our model captures the following features:
\begin{itemize}
\item Information is available at the transmitter at a random
time;
\item The transmitter chooses when to start sending
information;
\item Outside the information transmission period, the
transmitter stays  idle and the receiver observes
noise;
\item The receiver decodes without
knowing the information arrival time at the transmitter.
\end{itemize}


 Communication is discrete-time, and carried over a discrete memoryless channel
characterized by its finite input
and output alphabets $${\cal{X}} \cup\{\star\} \quad\text{and}\quad {\cal{Y}},$$
respectively, and transition probability matrix $$Q(y|x)\qquad x\in
{\cal{X}}\cup\{\star\},{ y \in{\cal{Y}}}.$$ Here $\star$ denotes the special
idle symbol, and $\cal{X}$ denotes the alphabet containing the symbols that can
be used in the actual transmission of the data. $\cal{X}$ may or may not contain
$\star$. We assume that no two different input symbols $x$ and $x'$ belonging to
${\cal{X}}$ have identical conditional distributions $Q(\cdot|x)$ and
$Q(\cdot|x')$.\footnote{This is without loss of generality, as two such symbols
are identical for communication purposes, so we can
consider the equivalent channel with one of these two symbols deleted from the
symbol alphabet.}
 \label{refQ}

Given $B$ information bits to be transmitted, a codebook ${\cal{C}}$ consists of
$2^B$ codewords of length $n$ composed of symbols from ${\cal{X}}$.
The message $m$ arrives at the transmitter at a random time $\nu$,
independent of $m$, and uniformly distributed over
$\{1,2,\ldots,A\}$, where the integer $A\geq 1$ characterizes the
{\emph{asynchronism
level}} between the transmitter and the receiver.  Only one message arrives
over the period $[1,2,\ldots,A+n-1]$.  If $A=1$, the
channel is said to be synchronous.

The transmitter chooses a time $\sigma(\nu,m)$ so that $$\nu \leq \sigma
(\nu,m)\leq A\qquad \text{almost surely}$$ to begin transmitting the codeword $c^n(m)\in
\cC$ assigned to message $m$.  This means that the transmitter cannot start
transmitting before the message arrives or after the end of the uncertainty
window. It turns out that the possibility to choose $\sigma$ as a function of
both $\nu$ and $m$ directly influences the cost to deliver this
information by allowing to convey information through timing.
In the rest of the paper, we suppress
the arguments $\nu$ and $m$ of $\sigma$ when these arguments are clear from
context. 

Before and after codeword transmission, {\it{i.e.}}, before time $\sigma$ and
after time $\sigma+n-1$, the receiver
observes ``pure noise.'' Specifically, conditioned on the event $\{\nu=t\}$,
$t\in \{1,2,\ldots,A\}$, and on the message to be conveyed $m$, the receiver
observes independent symbols $$Y_1,Y_2,\ldots,Y_{A+n-1}$$ distributed as
follows. For $$1\leq i\leq \sigma(t,m)-1$$ or $$\sigma(t,m)+n\leq i\leq
A+n-1\,,$$
the $Y_i$'s are distributed according to $ Q(\cdot|\star)$.  At any time $i\in
\{\sigma ,\sigma+1,\ldots,  \sigma+n-1\}$, the distribution is
$$Q(\cdot|{c_{i-\sigma+1}(m)})\,,$$ where $c_{i}(m)$ denotes the $i$\/th symbol
of the codeword $c^n(m)$.

\begin{figure}
\begin{center}
\input{canaltemps}
\caption{\label{grapheesss} Time representation of what is sent (upper arrow)
and what is received (lower arrow). The ``$\star$'' represents the ``idle''
symbol. Message $m$ arrives at time $\nu$, starts being sent at time $\sigma$,
and decoding occurs at time
$\tau$.}
\end{center}
\end{figure}
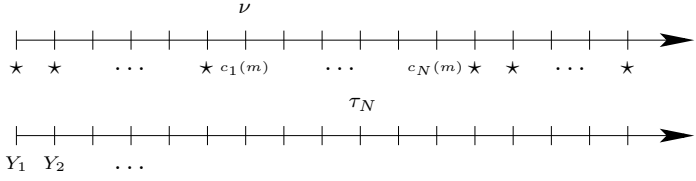

Knowing the asynchronism level $A$, but not the value of $\nu$, the receiver
decodes by means of a sequential test
$(\tau,\phi)$, where $\tau$ is a stopping time, bounded by $A+n-1$, with
respect to the output sequence $Y_1,Y_2,\ldots$ indicating when decoding
happens, and where $\phi$ denotes a decision rule that declares the decoded
message (see Fig.~\ref{grapheesss}). Recall that a (deterministic or
randomized) stopping time $\tau$ with respect to a sequence of random variables
$Y_1,Y_2,\ldots$ is a positive, integer-valued, random variable such that the
event $\{\tau=t\}$, conditioned on the
realization of $Y_1,Y_2,\ldots,Y_t$, is
independent of the realization of
$Y_{t+1},Y_{t+2},\ldots$, for all $t\geq 1$. Given $\{\tau=t\}$, $t\in
\{1,2,\ldots,A+n-1\}$, the function $\phi$ outputs a message based on the past
observations from time $1$ up to time
$t$.\footnote{To be more precise,  $\phi$ is any ${\cal{F}}_\tau$-measurable
function that takes values in the message set, where ${\cal{F}}_t$ is the
sigma field generated by  $Y_1,Y_2,\ldots, Y_t$.}

A ``code'' refers to a
codebook $\cC$ together with a decoder, {\it{i.e.}}, a
sequential test $(\tau,\phi)$.
Throughout the paper, whenever clear from context, we
often refer to a code using the codebook symbol $\cC$
only, leaving out an explicit reference to the decoder.

The maximum (over messages) decoding error probability
for a given code $\cC$ is defined as
\begin{align}\label{maxerror}
\pr({\EuScript{E}}|\cC)\triangleq\max_m\frac{1}{A}\sum_{t=1}^A
\pr_{m,t} ({\EuScript{E}}|\cC),
\end{align}
where the subscripts ``${m,t}$'' indicate conditioning on the
event that message $m$ arrives at
time $\nu=t$, and where $\EuScript{E}$ indicates the event that the decoded
message does not
correspond to the sent codeword, {\it{i.e.}},
$$\E\triangleq \{\phi(Y^\tau)\ne M\}$$
where $M$ denotes the random message to be transmitted.
\begin{defs}[Cost Function]
A  cost function $\ck: \cX\to [0,\infty)$ assigns a non-negative value
to each channel input.\footnote{``Kost'' is cost in German.}
\end{defs}

\begin{defs}[Cost of a Code]\label{costc}
The (maximum) cost of a code ${\cal C}$ is defined as
$$ \cK({\cal C})\triangleq \max_m \sum_{i=1}^n \ck(c_i(m)).$$
\end{defs}
\begin{defs}[Delay of a Code]\label{def:delaiscode}
Given $\varepsilon>0$, the (maximum) delay of a code ${\cal C}$, denoted by
$\cD({\cal
C}, \varepsilon)$, is defined as the smallest $d$ such that
$$\min_m\pr_m(\tau-\nu \leq  d-1) \geq
1-\varepsilon,$$
where $\pr_m$ denotes the output distribution conditioned on the sending of
message $m$.\footnote{Hence, by definition we have
$$\pr_m(\cdot)=\frac{1}{A}\sum_{t=1}^A\pr_{m,t}(\cdot)\,.$$}
\end{defs}
Throughout the paper, we often consider delays in the regime
$\varepsilon\to 0$. In this case, we omit an explicit reference to
$\varepsilon$. For instance, if $\{\cC_B\}$ is such that
$\cD(\cC_B,\varepsilon_B)=O(B)$ for some $\{\varepsilon_B\}$ such that
$\varepsilon_B\to 0$ as $B\to \infty$, we simply say that $\{\cC_B\}$ achieves a
delay that is linear in $B$---leaving implicit ``with probability asymptotically
equal to one.''

A key parameter we shall be concerned with is $$\beta\triangleq \frac{\log
A}{B}\,,$$ which we call the timing uncertainty per information bit.

Next, we define the asynchronous capacity per unit cost in the asymptotic regime
where $B\rightarrow \infty$ while $\beta$ is kept fixed.
\begin{defs}[Asynchronous Capacity per Unit Cost]
\label{def:cap}
\bm{R} is an achievable rate per unit cost at
timing uncertainty per information bit $\beta$ and delay exponent $\delta$ if
 there exists a sequence of codes $\{{\cal C}_B\}$, and a sequence of numbers
$\{\varepsilon_B\}$ with $\varepsilon_B\overset{B\to \infty}{\longrightarrow} 0$, such that
 $$\pr(\cE|{\cal{C}}_B)\leq \varepsilon_B\,,$$
$$\limsup_{B\rightarrow \infty} \log(\cD({\cal
C}_B,\varepsilon_B))/B \leq \delta\,,$$
and
$$ \liminf_{B \rightarrow \infty} \frac{B}{\cK({\cal C}_B)} \ge \bm{R}.$$
The asynchronous capacity per unit cost, denoted by $\bm{C}(\beta,\delta)$,
is the largest achievable rate per unit cost.
In the important case when $\delta =0$, we
define $\bm{C}(\beta) \triangleq
\bm{C}(\beta,0)$.
\end{defs}

Note that, in Definition~\ref{def:cap}, the codeword length $n$ is a
free parameter that can be optimized, just as for the synchronous
capacity per unit cost (see the comment after \cite[Definition
$2$]{Ve2}). The results in the next section characterize the
capacity per unit cost for arbitrary $\beta$ and $\delta$.  Similar to the
synchronous case, the results
simplify when there is a zero cost symbol, specifically
when $\cal{X}$ contains $\star$ and $\star$ has zero cost.

For simplicity, for the rest of the paper we assume that the only possible zero cost
symbol is $\star$---in particular, if $\star \notin \cX$ then $\cX$ contains
only non-zero cost symbols. The other, arguably unnatural,
cases can also be addressed by the arguments in this
paper and are briefly discussed in the remark before the proof of Theorem~\ref{gauss} in Section~\ref{sketches}.

\section{Results}\label{results}
Our first result gives the asynchronous capacity per
unit cost when $\delta=0$.
It can be viewed as the asynchronous analogue of Theorem 2 in
\cite{Ve2}, which states that the synchronous capacity per unit cost
is
\begin{align}\label{cpunitcost}
\max_{X}\frac{I(X;Y) }{\ex[\ck(X)]}
\,.
\end{align}
As mentioned above, in stating our results we assume that 
all non-$\star$ symbols in $\cX$ have positive cost, and that if
$\star$ is in $\cX$, then $\star$ has zero cost.
 \begin{thm}[Asynchronous Capacity per Unit Cost: Sub-exponential Delay
 Constraint]\label{ulimit}

The asynchronous capacity per unit cost at delay exponent $\delta=0$ is given by
\begin{equation}
\label{eq:main1}
\bm{C}(\beta) = \max_X \min\left\{\frac{I(X;Y)}{\ex[\ck(X)]} ,
\frac{I(X;Y ) + D(Y||Y_\star)}{\ex[\ck(X)](1 + \beta)}\right\},
\end{equation}
where $X$ denotes the random input to the channel,
$Y$ the corresponding output,  $Y_\star$ the random output of the channel when
the
idle symbol $\star$ is transmitted ({\it{i.e.}}, $Y_\star\sim Q(\cdot|\star)$),
$I(X;Y)$ the mutual information between $X$ and $Y$, and $D(Y||Y_\star)$ the
Kullback-Leibler distance between the distributions of $Y$
and~$Y_\star$.\footnote{$Y_\star$ is
interpreted as ``pure noise.''}

Furthermore, capacity
can be achieved by codes whose delay grows linearly in $B$.\footnote{See comment
after Definition~\ref{def:delaiscode}.}
\end{thm}

The two terms in \eqref{eq:main1} reflect the two constraints on
reliable communication. The first term corresponds to the standard
constraint that the number of bits that can reliably be
transmitted per
channel use  cannot exceed the input-output mutual information.
This constraint applies when the channel is synchronous,
hence also
in the absence of synchrony.

The second term in \eqref{eq:main1} corresponds to the receiver's ability to
determine the arrival time $\nu$ of the data. Indeed, even though the decoder
is only required to produce a message estimate, because of the delay constraint,
there is no loss in terms of capacity per unit cost to also require the decoder
to produce an approximate estimate of the time when transmission begins---the
delay
constraint implies that the decoder can locate the sent message within a time
window
that is negligible compared to $A$.  The quantity $$I(X;Y) +
D(Y||Y_\star)=D(XY||XY_\star),$$ where $D(XY||XY_\star)$ refers to the
Kullback-Leibler distance between the joint distribution of $(X,Y)$ and the
(product) distribution of $(X,Y_\star)$, measures how difficult it is for the
receiver to discern a data-carrying transmitted symbol from pure noise, and thus
determines how difficult it is for the receiver to get the timing correct.

When the alphabet $\cal{X}$ contains a zero-cost symbol $0$, the synchronous
result~\eqref{cpunitcost} simplifies, and Theorem 3 in \cite{Ve2} says that the
synchronous capacity per unit cost becomes
\begin{equation}
\max_{x\in
\cX}\frac{D(Y_x||Y_0)}{\ck(x)},
\label{eq:verdu}
\end{equation}
an optimization over the input alphabet instead of over the set of all input
distributions, where $Y_x$ refers to the output distribution given that $x$ is
transmitted.

We find an analogous simplification in the asynchronous setting
when $\star$ is in $\cX$ and has zero cost:


\begin{thm}[Asynchronous Capacity per Unit Cost With Zero Cost Symbol:
Sub-exponential Delay Constraint]\label{ulimit2} If $\star$ is in
$\cal{X}$ and has zero cost, the asynchronous
 capacity per unit cost at delay exponent $\delta=0$ is given by
\begin{equation}
\label{eq:main2}
\bm{C}(\beta) = \frac{1}{1+\beta} \max_{x\in
\cX}\frac{D(Y_x||Y_\star)}{\ck(x)}\,,
\end{equation} and capacity
can be achieved by codes whose delay grows linearly with $B$.
\end{thm}
Hence, a lack of synchronization multiplies the cost
of sending one bit of information by $1+\beta$.
An intuitive justification for this is as follows. Suppose there exists an
optimal coding scheme
that can both isolate and locate the sent message with high
probability---as alluded to above, the ability to ``locate''
the message is a consequence of the decoder's delay
constraint. Assuming that the delay is negligible, {\it{i.e.}},
the delay grows subexponentially with $B$, this
allows us to consider message/location pairs as inducing a code of
size $$ \approx 2^B A$$ used for communication across the
{\emph{synchronous channel}}. Hence, since $A=2^{\beta B}$ we are effectively communicating
$$\approx \beta B+B=B(1+\beta)$$ bits reliably over the synchronous
channel. Therefore, sending $B$ bits of information at
asynchronism level $\beta$ is at least as costly as sending
$B(1+\beta)$ bits over the synchronous channel. Flipping this reasoning
around, the asynchronous channel effectively induces a codebook for
message/location pairs where the location is encoded via {\it pulse position
modulation} (PPM).
From \cite{Ve2},  optimal coding schemes are similar to PPM
in that the codewords consist almost entirely of the
zero cost symbol. This provides an intuitive
justification for why $(1+\beta)\ck_{\rm sync}$ is an achievable
rate per unit cost.

Theorem \ref{ulimit2} can be extended to the (continuous-valued)
Gaussian channel, where the idle symbol
$\star$ is the $0$-symbol:
\begin{thm}[Asynchronous Capacity per Unit
Cost for the Gaussian Channel: Sub-exponential Delay Constraint]\label{gauss}
The asynchronous
 capacity per unit cost for the Gaussian
 channel with variance $N_0/2$, quadratic cost function ({\it{i.e.}},
$k(x)=x^2$), and delay exponent $\delta=0$, is given by
\begin{align}\label{zot}\C(\beta)= \frac{1}{1+\beta} \frac{\log e}{N_0}, \quad
\beta \geq 0\,.\end{align}
 \end{thm}

Theorem \ref{ulimit} can be extended to the case of a large delay constraint,
{\it{i.e.}}, when $0<\delta\leq\beta$.
In this case, the formula for capacity is slightly
different depending on whether $\star$ is in $\cal{X}$ or not, 
as stated in the following result.

 \begin{thm}[Asynchronous Capacity per Unit Cost: Exponential Delay
Constraint]\label{ulimit3}
The asynchronous
 capacity per unit cost at delay constraint $\delta$, with $0\leq \delta\leq\beta$,
is given by:
\begin{itemize}
\item[(a)] if $\star \in \cX$ and $\star$ has zero cost then
\begin{equation}\nonumber
\bm{C}(\beta,\delta) = \bm{C}(\beta-\delta),
\end{equation}
{\it{i.e.}}, it is the same as the capacity per unit cost with delay exponent
$\delta=0$, but with asynchronism exponent $\beta$ reduced to $\beta-\delta$;
\item[(b)] if $\star$ is not in $\cX$ and all non-$\star$ symbols
have positive cost then
\begin{align}
&\bm{C}(\beta,\delta)\nonumber \\
& = \max_X \min\left\{\frac{I(X;Y)}{\ex[\ck(X)](1-\delta)} ,
\frac{I(X;Y ) + D(Y||Y_\star)}{\ex[\ck(X)](1 + \beta-\delta)}\right\}.
\end{align}
\end{itemize}
\end{thm}

The uniform distribution on $\nu$ in the model is not critical. The
next result extends Theorem~\ref{ulimit} to the case where $\nu$ is
non-uniform. For a non-uniform distribution on $\nu$, what is
important turns out to be its ``smallest'' set of mass
points that contains ``most'' of the probability.

Consider a general arrival time ${\nu}$ (defined over the positive
integers), not necessarily bounded. For a given $\varepsilon>0$, let
${\cal{S}}(\varepsilon)$ denote the smallest subset of the support of ${\nu}$
({\it{i.e.}}, the set of $n$ such that $\pr(\nu=n)>0$)  whose probability is at
least
$1-\varepsilon$. Hence,  $\pr(\nu\in {\cal{S}}(\varepsilon))\geq
1-\varepsilon$ by definition.

\begin{thm}[Asynchronous Capacity per Unit Cost With
Non-uniform Arrival Time:
Sub-exponential Delay Constraint]\label{ulimit4}
For a given sequence of arrival times $\{{\nu}_B\}_{B\geq 1}$, define
\begin{equation}\label{def:lim}
\bar{\beta} = \inf_{\{\varepsilon_B\}} \limsup_{B\rightarrow \infty}
\frac{\log(|{\cal{S}}(\varepsilon_B)|)}{B},
\end{equation}
where the infimum is  with respect to all sequences $\{\varepsilon_B\}$
of nonnegative numbers such that $\lim_{B\rightarrow
\infty} \varepsilon_B=0$.

Then, the asynchronous
 capacity per unit cost at delay exponent $0$ is given by
\begin{equation*}
\bm{C}(\bar{\beta}) = \max_X \min\left\{\frac{I(X;Y)}{\ex[\ck(X)]} ,
\frac{I(X;Y) + D(Y ||{Y_\star})}{\ex[\ck(X)](1 +
\bar{\beta})}\right\}\,.
\end{equation*}
\end{thm}
Although the formula for $\bar{\beta}$ in \eqref{def:lim} appears
unwieldy, in many cases it can easily be evaluated. For example, in
many cases, such as the uniform or geometric distributions, the
formula reduces to the normalized entropy
$$\bar{\beta} = \lim_{B\to \infty} H(\nu_B)/B\,.$$
There are cases, however, where \eqref{def:lim} doesn't
reduce to the normalized entropy. For instance, consider the
case when $\nu_B=1$ with probability $1/2$,
and $\nu_B=i$ with probability $(1/2)2^{-\beta B}$
for $i=2,\ldots,2^{\beta B}+1$. Then, $\bar{\beta}=\beta$ and
$H(\nu_B)=1+0.5\beta B$, which yields $$\bar{\beta}= 2
\lim_{B\to \infty} H(\nu_B)/B\,.$$

\subsection*{Asynchronous Capacity}
\vspace{0.06in}

The above results focus on characterizing the asynchronous capacity {\em per
unit cost}. However, just as the synchronous capacity per unit cost result
(\ref{cpunitcost}) immediately implies the standard (synchronous) capacity
result\footnote{Information per
symbol and information per unit cost are differentiated by lightface
and boldface characters, respectively, as in \cite{Ve2}.}
$$ C = \max_X I(X;Y)$$
by setting the cost function $k(\cdot)=1$, Theorem~\ref{ulimit} implies
the asynchronous capacity result
\begin{align}\label{capacitybeta}C(\beta)=\max_X \min
\left\{I(X;Y);\frac{I(X;Y)+D(Y||Y_\star)}{1+\beta}\right\}\,,
\end{align}
the largest number of
information bits per {\it transmitted symbol} that can be supported reliably by
an asynchronous
channel, as a function of $\beta$.

Instead of $\beta$, we may alternatively consider the asynchronism parameter
$\alpha=(\log A)/n=\beta R$ introduced in \cite{CTW,TCW}. Using
\eqref{capacitybeta}, we deduce that rate $R$
 is achievable if and only if, for some
input $X$,
$$R\leq I(X;Y)$$
and
$$R\leq D(XY||XY_\star)-\alpha\,.$$
Hence, asynchronous capacity is alternatively given by
\begin{align}\label{mx1}C(\alpha)=\max&\Big\{\max_{X:D(Y||Y_\star)\geq
\alpha}I(X;Y);\nonumber\\
&\max_{X: D(Y||Y_\star)\leq
\alpha}D(XY||XY_\star)-\alpha\Big\}\,,
\end{align}
with the convention that the maximum evaluates to $0$ if the set being
optimized over is empty.
Consider the second inner maximization in \eqref{mx1}. Since $D(XY||XY_\star)$ is convex in $X$, and
the set $\{X: D(Y||Y_\star)\leq
\alpha\}$ is convex, the maximum is achieved for some
extreme point of the set, {\it{i.e.}}, either for some $X$ such that
$D(Y||Y_\star)=\alpha$, or for a distribution $X$ concentrated on a single point
and such that $D(Y||Y_\star)<\alpha$.
However, in the latter case we have $$D(XY||XY_\star)-\alpha < 0$$
since $D(XY||XY_\star)=D(Y||Y_\star)< \alpha$.
Thus,~\eqref{mx1} reduces to
\begin{align*}C(\alpha)=&\max_{X:D(Y||Y_\star)\geq
\alpha}I(X;Y)\,.
\end{align*}

Although not explicit in the statement of
Theorem~\ref{ulimit}, the proof of this theorem shows that
$C(\alpha)$ can be achieved with codes whose delays  are no larger than $n$.
Summarizing the above discussion, we get:
\begin{cor}
The capacity at delay exponent $\delta=0$, and with respect to asynchronism
parameter
$\alpha=(\log A)/n$, is given by
\begin{align*}C(\alpha)=&\max_{X:D(Y||Y_\star)\geq
\alpha}I(X;Y)\,.
\end{align*}
Furthermore, capacity is achievable with codes whose delays are no larger than
$n$.
\end{cor}
A closely related problem is determining the capacity when rate is defined in
terms of bits
per {\it received symbol}. For this problem, we refer the reader to
\cite{TCW,TCWj2}, where  capacity as a function of $\alpha$
is studied, and where
rate is defined with respect to
the expected elapsed time between the instant information is available at the
transmitter
and the instant it is decoded.

\section{Proofs of Results}\label{sketches}
We use $\cP^\cX$ to denote the set of
distributions over the finite alphabet $\cX$.
Recall that the type of a string
$x^n\in \cX^n$, denoted by $\hat{P}_{x^n}$, is the probability distribution over
$\cX$ that
assigns, to each $a\in \cX$, the number of occurrences of $a$ within $x^n$
divided by $n$~\cite[Chapter $1.2$]{CK}. For instance, if $x^3=010$, then
$\hat{P}_{x^3}(0)=2/3$ and $\hat{P}_{x^3}(1)=1/3$. The joint type
$\hat{P}_{x^n,y^n}$ induced by a pair of strings $(x^n,y^n)\in \cX^n\times
\cY^n$ is
defined similarly. The set of strings of length $n$ that have type $P$ is
denoted by $\cT_P$, and is called the ``type class of $P$.'' The set of all
types over $\cX$ of strings of length $n$ is
denoted by $\cP_n^{\cX}$.

Given a string $x^n\in \cX^n$ and a conditional probability distribution
$W=\{W(y |x)$, $(x,y)\in \cX\times \cY\}$, the
set of strings $y^n$ that have conditional type $W$ given $x^n$ is denoted
by $\cT_W(x^n)$, {\it{i.e.}},
$$\cT_W(x^n)\triangleq\{y^n\in \cY^n:\hat{P}_{x^n,y^n}=\hat{P}_{x^n}W\}\,.$$

 Finally, we use the standard ``big-O'' Landau notation to characterize growth
rates
(see, e.g., \cite[Chapter~3]{CLRS}), and use $\poly(\cdot)$ to denote a function
that does not grow
or decay faster than polynomially in its argument.

The following two standard results on types are often used in the analysis:
\begin{fact}[\hspace{-.01cm}{\cite[Lemma~2.2]{CK}}]
\label{fact:1}
\begin{align*}
|\cP_n^{\cX}| &=\poly(n)\,.
\end{align*}
\end{fact}

\begin{fact}[\hspace{-.01cm}{\protect\cite[Lemma 2.6]{CK}}]
\label{fact:2}
If $X^n$ is independent and identically distributed (i.i.d.) according
to $X_1\sim P_1$, then
\begin{equation*}
\poly(n)e^{-nD(X_2\|X_1)}\leq \pr(X^n\in \cT_{P_2}) \leq  e^{-nD(X_2\|X_1)}
\end{equation*}
for any $X_2\sim P_2\in\cP^\cX_n$.
\end{fact}

\begin{IEEEproof}[Achievability of Theorem~\ref{ulimit}]
We first show the existence of a random code that achieves the asynchronous
capacity per unit cost when the latter is computed with respect to average error
probability.
A standard expurgation argument then shows the existence of a deterministic code
achieving the same (asymptotic) performance as the random code, but now with
respect to maximum error probability.

Fix some arbitrary distribution $P$ on $\cal{X}$. Let $X$ be the input having
that distribution,
and let $Y$ be the corresponding output, {\it{i.e.}}, $(X,Y)\sim
P(\cdot)Q(\cdot|\cdot)$.

Given $B$ bits of information to be transmitted, the codebook ${\cal C}$ is
randomly
generated as follows.
For each message $m\in \{1,2,\ldots, 2^B\}$, randomly generate a
length $n$ sequence $x^n$ i.i.d. according to $P$. If $x^n$ belongs to the
``constant composition'' set\footnote{$||\cdot ||$ refers to the $L_1$-norm.}
\begin{align}\label{constprop}
{\cal{A}}=\{x^n:||\hat{P}_{x^n}-P||\leq 1/\log n\}\,,
\end{align}
we let $c^n(m)=x^n$. Otherwise, we repeat the procedure until we generate a
sequence sufficiently close to $P$. From Chebyshev's inequality, for a fixed
$m$, it is very unlikely that any repetition will be required to generate
$c^n(m)$,
{\it{i.e.}},
\begin{align}\label{typset}
P^n({\cal{A}})\to 1\quad \text{as}\quad n\to \infty,
\end{align}
where $P^n$ denotes the order $n$ product distribution of $P$.

The obtained codebook is thus essentially of constant composition,
{\it{i.e.}}, each symbol appears roughly the same number of times across
codewords. Moreover, by construction all codewords in the random ensemble have
cost $$n\ex[k(X)](1+o(1))$$ as $n\rightarrow \infty$.

The sequential typicality decoder operates as
follows.  At time $t$, for each
$m\in\{1,2,\ldots,2^B\}$, it
computes the empirical distributions
$$\hat{P}_{c^n(m),y^{t}_{t-n+1}}(\cdot,\cdot)$$
induced by $c^n(m)$ and the $n$ output symbols
$y_{t-n+1}^t$. If there is a unique message $m$ for
which
$$||\hat{P}_{c^n(m),y^{t}_{t-n+1}}(\cdot,\cdot)-P(\cdot)Q(\cdot|\cdot)||\leq
2/\log n,$$ the decoder stops and declares that
message $m$ was sent. If more than one codeword is
typical, the decoder stops and declares one of the
corresponding messages uniformly at random.\footnote{The
notion of typicality we use is often referred to as
``strong typicality'' in the literature.} If no
codeword is typical at time $t$,  the decoder moves
one step ahead and repeats the procedure based on
$Y^{t+1}_{t-n+2}$. If the decoder reaches time $A+n-1$ and no codeword is
typical, then it declares a randomly and uniformly chosen message.

We first compute the error probability averaged over codebooks and messages.
Suppose message $m$ is  transmitted. The error event that the decoder declares
some specific message $m'\ne m$ can be decomposed as\footnote{Notice that the
decoder outputs a message with probability one by time $A+n-1$.}
\begin{align}\label{decom}
\{m\to m'\}=\E_1\cup\E_2\,,
\end{align}
where the error
events $\E_1$ and $\E_2$ are defined as
\begin{itemize}
\item $\E_1$: the decoder stops at a time $t$ between $\nu$ and
$\nu + 2n-2$  (including $\nu$ and $\nu+2n-2$), and declares $m'$;
\item $\E_2$: the decoder stops
either at
a time $t$ before $\nu$ or from $\nu
+ 2n-1$ onwards, and declares $m'$.
\end{itemize}
For the error event $\E_1$, for some $0 \leq
k \leq n-1$ the first or the last $k$ symbols of
$Y^n$ are generated by noise, and the remaining $n-k$
symbols are generated by the sent codeword $C^n(m)$.\footnote{We use a capital
letter for $C^n(m)$ since codewords are
randomly generated.} The
probability that such a $Y^n$ together with $C^n(m')$
yields an empirical distribution $J$ that is jointly typical with
$P(\cdot)Q(\cdot|\cdot)$, that is,
\begin{align}\label{close}
||J(\cdot,\cdot)-P(\cdot)Q(\cdot|\cdot)||\leq
2/ \log n\,,
\end{align}
is upper bounded as
\begin{align}\label{mezzomezzo}
&\pr_m(\hat{P}_{C^n(m'),Y^n}=J)\nonumber\\
&=\sum_{y^n\in {\cal{Y}}^n}\pr_m(Y^n=y^n)\sum_{x^n :
\hat{P}_{x^n,y^n}=J} \pr_m(X^n=x^n)\nonumber \\
&\leq \sum_{y^n\in {\cal{Y}}^n}\pr_m(Y^n=y^n)\sum_{x^n:
\hat{P}_{x^n,y^n}=J} 2^{-n(H(J_\cX)+D(J_\cX||P)-\varepsilon)}\nonumber \\
&\leq \sum_{y^n\in {\cal{Y}}^n}\pr_m(Y^n=y^n) 2^{-n(
H(J_\cX)-\varepsilon)}|\{x^n:
\hat{P}_{x^n,y^n}=J\}|\nonumber \\
&\leq \sum_{y^n\in {\cal{Y}}^n}\pr_m(Y^n=y^n) 2^{-n(
H(J_\cX)-\varepsilon)}2^{n H
(J_{\cX|\cY})}\nonumber \\
&\leq   2^{-n (I(J)-\varepsilon)}\nonumber \\
&\leq 2^{-n (I(X;Y)-2\varepsilon)}\end{align}
for any $\varepsilon>0$ and all $n$ large enough,
 where $H(J_\cX)$ denotes the entropy of the left marginal of
$J$, where $$H(J_{\cX|\cY})\triangleq -\sum_{b\in \cY}
J_\cY(b)\sum_{a\in \cX} J_{\cX|\cY}(a|b)\log
J_{\cX|\cY}(a|b),$$ and where $I(J)$ denotes the mutual information induced
by~$J$.

The first equality in \eqref{mezzomezzo} follows from the independence of
$C^n(m')$ and $Y^n$, since $Y^n$ corresponds to the output of $C^n(m)$. For the
first inequality, note that if the codewords were randomly generated
with each component of each codeword i.i.d. according to $P$, we could
deduce
from \cite[Theorem $11.1.2$, p. 349]{CT} that
$$ P^n(X^n=x^n)= 2^{-n(H(J_\cX)+D(J_\cX||P))}\,.$$
The actual (non-i.i.d) codeword distribution is the i.i.d. distribution,
conditioned
on the constant composition event \eqref{constprop}. Therefore, we have
$$\pr_m(X^n=x^n)=\left\{\begin{array}{cc}
\frac{P^n(X^n=x^n)}{P^n({\cal{A}})})& x^n \in
{\cal{A}}\\
0 & \text{otherwise,}
\end{array}
\right.$$
and from \eqref{typset} we get
$${\pr}_m(X^n=x^n)= 2^{-n(H(J_\cX)+D(J_\cX||P))}(1+o(1))$$
as $n\rightarrow \infty$, uniformly over the set ${\cal{A}}$. This justifies the
first inequality in \eqref{mezzomezzo}. The second inequality in
\eqref{mezzomezzo}
follows from the non-negativity of the Kullback-Leibler distance.
The third
inequality in \eqref{mezzomezzo} follows from \cite[Lemma $2.5$, p. $31$]{CK}.
The fourth inequality holds since $H(J_\cX)-H(J_{\cX|\cY})=I(J)$, and by
upperbounding the sum of the probabilities by one. Finally, the fifth
inequality in  \eqref{mezzomezzo}
holds for any $\varepsilon>0$ and all $n$ large
enough since, by assumption, $J$ is close to
$PQ$ (see \eqref{close}).

From \eqref{mezzomezzo}, by taking a union bound over all empirical
distributions
$J$ that are jointly typical with $PQ$ ($\poly(n)$ by Fact~\ref{fact:1}) and
over all the (less than
$2n$) times involved in $\E_1$, we
obtain the upper bound
\begin{align}\label{e1}
\pr_m(\E_1) \leq
2^{-n(I(X;Y)-3\varepsilon)}
\end{align}
for all $n$ large enough.

For the second error event $\E_2$, pure noise produces some
output $Y^n$ that is jointly typical with
$C^n(m')$. The probability that a noise generated $Y^n$ together with $C^n(m')$
yields an empirical type $J$ is upper bounded by
$$2^{-nD(J||XY_\star)}$$
by \cite[Lemma 1.2.6]{CK}---recall that $D(J||XY_\star)$ refers to the
Kullback-Leibler
distance between, on the one hand, the joint distribution $J$, and on
the other hand, the product of the
distributions of $X$ and $Y_\star$. Hence, by taking a union bound over all
typical $J$'s that
satisfy \eqref{close} ($\poly(n)$ of them by Fact~\ref{fact:1}), and by
using the continuity of the Kullback-Leibler
distance,
\footnote{Technically, the divergence is not continuous if,
for example, both distributions are $0$ at the same point. However,
at points of discontinuity, the discontinuity can only help since the divergence becomes infinite, and it is easily seen
that the  corresponding error event has zero probability.
\label{footnote:cont}} 
the probability that a noise
generated $Y^n$ is typical with $C^n(m')$ is upper bounded
by
$$2^{-n(D(XY||XY_\star)-\varepsilon)}=2^{-n(I(X;Y)+D(Y||Y_\star)-\varepsilon)}\,,$$
for any $\varepsilon>0$ and all $n$ large enough.
 Finally, by taking a union bound over all (less than $A$) times where
noise could produce such an output, we get
\begin{align}\label{e2}\pr_m(\E_2) \leq  A\cdot
2^{-n(I(X;Y)+D(Y||Y_\star)-\varepsilon)},\end{align}
for any $\varepsilon>0$ and all $n$ large enough.

Combining \eqref{decom}, \eqref{e1}, and \eqref{e2}, we get
\begin{align*}\pr_m(m \rightarrow m')&=\pr_m(\E_1)+\pr_m(\E_2)\nonumber \\
&\leq
2^{-n(I(X;Y)-3\varepsilon)} \\
&+ A \cdot 2^{-n(I(X;Y) +
D(Y||Y_\star)-\varepsilon)},\end{align*}
for any $\varepsilon>0$ and all $n$ large enough.

Hence, by taking a union bound over all possible wrong
messages, we obtain that for any $\varepsilon>0$,
\begin{align*}\pr_m({\E})\leq 2^B\Big( &
2^{-n(I(X;Y)-3\varepsilon)} \\
&+ A \cdot 2^{-n(I(X;Y) +
D(Y||Y_\star)-\varepsilon)}\Big)\,,\end{align*}
for $n$  large enough and all $m$.  Since
the above bound is valid for a randomly generated
code, we deduce that
\begin{align}\label{er11}\ex_{\cal{C}}(\bar{\pr}({\E}|{\cal{C}}))&=\pr_m({\E})\nonumber\\
&\leq 2^B\Big(
2^{-n(I(X;Y)-3\varepsilon)}\nonumber\\
& \hspace{.3cm}+ A \cdot 2^{-n(I(X;Y) +
D(Y||Y_\star)-\varepsilon)}\Big)\nonumber\\
& \triangleq \varepsilon_1(n),\end{align}
where $\bar{\pr}({\E}|{\cal{C}})$ denotes the error probability of code
${\cal{C}}$ averaged over the messages.

We now turn to the delay of the code. Suppose message $m$ is  transmitted
with a specific (non-random) codeword $c^n(m)$ that belongs to the set
$\cal{A}$.
If event $$\{\tau\geq \nu +n\}$$ happens, then necessarily
$Y^{\nu+n-1}_\nu$ isn't typical with $c^n(m)$. By Chebyshev's
inequality, the
probability of the latter event tends to zero as $n\rightarrow
\infty$, hence $$\pr_m(\tau\leq \nu+n)\geq 1-\varepsilon_2(n),$$ where
$\varepsilon_2(n)$ is a function that tends to zero as
$n\rightarrow \infty$. Since the above inequality holds for any specific
codeword that belongs to $\cA$, we get
\begin{align}\label{deln}
\cD(\cC,\varepsilon_2(n))\leq n
\end{align}
for any code $\cC$ whose codewords belong to $\cA$.

The proof can now be concluded. From inequality  \eqref{er11}, there exists a
specific code $\cC\subset{\cA}$ whose error
probability, averaged over messages, is less than
$\varepsilon_1(n)$.  Removing the half of the codewords with the highest error
probability, we end up with a set ${\cal{C}}'$ of $2^{B-1}$ codewords
whose maximum error probability $\pr(\E)$ satisfies
\begin{align}\label{mwi}\pr(\E)\leq
2\varepsilon_1(n)\,,
\end{align} and whose delay satisfies
$$\cD(\cC',\varepsilon_2(n))\leq n$$
by the previous argument.

Now, fix the ratio $B/n$, thereby imposing a delay linear
in $B$, and substitute $A = 2^{\beta B}$ in the
definition of $\varepsilon_1(n)$ (see \eqref{er11}). Then, $\pr(\E)$ goes to
zero as $B\rightarrow  \infty$ whenever
\begin{align}
\label{cn1}
\frac{B}{n} < \min\bigg\{I(X;Y),
\frac{I(X;Y)+D(Y||{Y_\star})}{1+\beta}\bigg\}.
\end{align}
Recall that,  by construction, all the codewords  have cost $n\ex[k(X)](1+o(1))$
as
$n\rightarrow \infty$. Hence, for any $\eta>0$ and all $n$ large enough,
\begin{align}\label{kos}k(\cC')\leq n\ex[k(X)](1+\eta)\,.
\end{align}
Condition \eqref{cn1} is thus implied by
condition \begin{align}\label{eqmaster}
\frac{B}{\cK({\cal C}')} <
\min\bigg\{\frac{I(X;Y)}{(1+\eta)\ex[\ck(X)]},
\frac{I(X;Y)+D(Y||{Y_\star})}{\ex[\ck(X)](1+\eta)(1+\beta)}\bigg\}.
\end{align}
 Maximizing over all input distributions, and using the
 fact that $\eta>0$ can be chosen arbitrarily, proves that the right-hand side
of~\eqref{eq:main1} is asymptotically achieved by
non-random codes with delay at most $n$, which grows linearly
with~$B$.
\end{IEEEproof}
\begin{rem} From \eqref{eqmaster} it follows that whenever there
exists some input $X$ such
that $I(X;Y)>0$ while $\ex [k(X)]=0$, and thus $\cal{X}$ contains more than one
zero cost symbol, the asynchronous capacity per unit cost is infinite,
{\it{i.e.}, $\bm{C}(\beta)=\infty$, for any $\beta\geq 0$}.
\end{rem}

\begin{IEEEproof}[Achievability of Theorem~\ref{ulimit3}]
The achievability scheme for Theorem~\ref{ulimit3} is similar to the
achievability scheme used to prove Theorem \ref{ulimit} except that we
distinguish the cases $\star \in \cal{X}$ and $\star \notin \cal{X}$.

\noindent {\emph{(a) $\star \in \cal{X}$:}}
 The main change is
that now the transmitter does not start transmitting at time
$\nu$. Instead, the transmitter only starts transmitting
at the first multiple of $2^{\delta B}$ larger than $\nu$, so that now
$\sigma$ takes values over multiples of $2^{\delta B}$. Such a
transmission scheme reduces
the receiver's uncertainty about $\sigma$ from
uniformly over $2^{\beta B}$ time slots to (essentially) uniformly
over only $2^{(\beta-\delta) B}$ time slots.

One proves that $\bm{C}(\beta-\delta)$ is achievable with delay
$O(2^{\delta B})$ by repeating the arguments for the achievability of
Theorem~\ref{ulimit}. The random codebook is constructed so that each codeword
satisfies the constant
composition property. The blocklength $n$ is still chosen
to be $O(B)$ so that, in contrast with the achievability of
Theorem~\ref{ulimit}, where delay and blocklength are the same, now
the blocklength is exponentially smaller than the delay.

The rest of the analysis is essentially unchanged. Since the
codewords are constructed in the same way, the cost is unchanged,
and the probability of error analysis is the same, except that $A$ is
replaced by ${A}/{2^{\delta B}}$ because now the
transmission timing allows the decoder to only consider
${A}/{2^{\delta B}}$ time slots instead of all $A$ time slots.
Therefore, $\beta$ is replaced by $\beta-\delta$, completing the
proof.

\noindent {\emph{(b) $\star \not\in \cal{X}$:}}
The main change is that the transmitter uses the freedom
in the choice of $\sigma$ to communicate part of the information through timing;
$B(1-\delta)$ information bits are contained in each codeword and $B\delta$
information bits are conveyed via timing. To achieve this, we use a space-time code.

The transmitter generates $2^{B(1 - \delta)}$ random codewords in the same way 
as in the achievability proof of Theorem~\ref{ulimit} to obtain a codebook $$\{c^n(s)\quad \text{with}\quad 1\leq s\leq 2^{(1-\delta)B}\}\,.$$ 
Label each of the $2^{B}$ messages with one of the $2^{(1-\delta)B}\times 2^{\delta B}$ pairs of integer indices $(s,j)$, {\it{i.e.}}, the message set is given by 
$$\{m(s,j)\quad\text{with}\quad 1\leq s\leq 2^{(1-\delta)B},  1\leq j\leq 2^{\delta B} \}\,.$$
(For simplicity we assume that $2^{B(1-\delta)}$ and $2^{\delta B}$ are integers.) 
For any (space) index $s\in \{1,2,\ldots,2^{(1-\delta) B}\}$, the set of messages  
$$\{m(s,j), 1\leq j\leq 2^{\delta B} \}$$
is associated to codeword $c^n(s)$.

Transmission always starts at a time that is a multiple of $n$. 
Suppose message $m$ arrives at time $\nu$ and that $m=m(\bar{s},\bar{j})$.
The transmitter first computes the ``offset'' 
$$O = \bar{j}-\left\lceil \frac{\nu}{n}\right\rceil \mbox{ mod } 2^{\delta B}.$$
The transmitter then starts sending codeword $c^n(\bar{s})$ 
at time 
\begin{align}\label{start}
\sigma(\nu,m)=\left(\left\lceil \frac{\nu}{n}\right\rceil+O\right)n.
\end{align}

The receiver uses a sequential typicality decoder to find the transmitted codeword as in the proof of the achievability part of Theorem~\ref{ulimit}---since transmission times are restricted to be multiples of $n$, the sequential typicality decoder can be restricted to multiples of~$n$. 

Suppose codeword $\hat{s}$ is found to be typical at time $t$. The receiver then computes the estimate $\hat{\sigma}$ for $\sigma$ given by 
 $$\hat{\sigma}=t-n+1$$ and finds the index $\hat{j}\in \{1,2,\ldots,2^{(1-\delta)B}\}$ such that
 $$\hat{j}=\frac{\hat{\sigma}}{n}\mod 2^{\delta B}\,.$$
The receiver then declares $\hat{m}=m(\hat{s},\hat{j})$.

The rest of the analysis is essentially unchanged. Since the
codewords are constructed in the same way, the cost is unchanged,
and the probability of error analysis is the same, except that $2^B$ is
replaced by $2^{B(1-\delta)}$ because the
transmission timing allows the decoder to only consider
$2^{B(1-\delta)} $ codewords instead of $2^B$ codewords. 
\end{IEEEproof}

\begin{IEEEproof}[Achievability of Theorem~\ref{ulimit4}]
To prove the achievability part of Theorem~\ref{ulimit4}, one applies
essentially the same
arguments as for the achievability of Theorem~\ref{ulimit}. The transmitter's
strategy is unchanged, {\it{i.e.}}, $\sigma=\nu$, and a random codebook
satisfying
the constant composition property is used to encode the messages.
At the receiver, we need a suitable analog of the set $\{1,2,\ldots,A\}$
of time slots to consider. A natural choice is to pick a sequence of nonnegative
numbers $\{\varepsilon_B\}$ such that $\varepsilon_B\overset{B\to
\infty}{\longrightarrow} 0$,  and, for each $B$, consider the ``typical'' set
${\cal{S}}(\varepsilon_B)$ whose probability, under the arrival time
distribution, is at least $1-\varepsilon_B$ by definition.
The receiver operates just as before, {\it{i.e.}}, using a sequential typicality
decoder,
but only over the set of times in~${\cal{S}}(\varepsilon_B)$.

Since the codewords are constructed in the same way, the cost of the
codebook is unchanged. The probability of error and delay analysis
now breaks into two cases: $\nu\in {\cal{S}}(\varepsilon_B)$ and
$\nu\notin {\cal{S}}(\varepsilon_B)$. The case $\nu \in
{\cal{S}}(\varepsilon_B)$ is handled as previously, except that $A$ is replaced
by
$|{\cal{S}}(\varepsilon_B)|$. When $\nu \not \in
{\cal{S}}(\varepsilon_B)$, we make the
worst-case assumption that the message is wrongly decoded and that
the delay is infinite. We can afford to do this because
$\pr(\nu \not \in {\cal{S}}(\varepsilon_B))\overset{B\to
\infty}{\longrightarrow}0$ by definition. Hence, the event $\{ \nu \not \in
{\cal{S}}(\varepsilon_B)\}$ has a vanishing effect on the probability of error
and
the delay. Optimizing over the choice of sequence
$\{\varepsilon_B\}$ completes the proof.
 \end{IEEEproof}

\begin{IEEEproof}[Converses of Theorems~\ref{ulimit} and~\ref{ulimit3}]

Assume that $\{{\cal{C}}_B\}$ achieves a rate per unit cost $\R>0$ at
timing uncertainty per information bit $\beta$ and delay exponent $\delta$ with
$0\leq \delta\leq \beta$. Recall that the delay constraint means that
\begin{align}\label{delcs}
\limsup_{B\rightarrow
\infty}\frac{\log \cD_B(\cC_B,\varepsilon_B)}{B}=\delta
\end{align}for some
sequence of non-negative numbers $\varepsilon_B\to 0$ as $B\to
\infty$.  
To establish the converses, we use the following concept of ``extended
codewords.''
To shorten notation, for the rest of the proof we use $\cD_B$ instead of
$\cD(\cC_B,\varepsilon_B)$.

\noindent {\emph{Extended codewords:}} An extended codeword
for a given message $m$ consists of the sequence of symbols that are transmitted
from time $\nu$ until time $\nu+\cD_B-1$. Hence, for
$\nu+d_B-1\geq
\sigma+n$, the
codeword corresponding to message $m$ consists of $\star$'s from time
$\nu$ until time $\sigma-1$, followed by $c^n(m)$, followed
by $\star$'s until time $\nu+d_B-1$.  Instead,
if $\nu+d_B\leq \sigma+n$, the codeword
corresponding to message $m$ consists of $\star$'s from time $\nu$
until time $\sigma-1$, followed by the first
$\nu+d_B-\sigma$ symbols of
$c^n(m)$. The cost of  the extended codeword, which we simply denote by
$c(m)$, is defined to be the same as the cost of $c^n(m)$.

From now on, codewords always refer
to extended codewords, and codebooks always refer to
sets of extended codewords.

To establish the theorems, we show that for any $\eta>0$ and all $B$ large
enough, $\R$ and
$\beta$ satisfy
\begin{align}\label{ratei}
\R \ex[\ck(X)]\leq I(X;Y)(1+\eta)
\end{align}
if $\star \in \cal{X}$ and $\star$ has zero cost, or
\begin{align}\label{ratei2}
\R \ex[\ck(X)]\leq \frac{I(X;Y)}{1-\delta}(1+\eta)
\end{align}
if $\star \not \in \cal{X}$ and all non-$\star$ symbols have
positive cost.
In either case, we also show that
\begin{align}\label{divi}
\R \ex[\ck(X)] (1+\beta-\delta-\eta) \leq
D(XY||XY_\star),
\end{align}
where $X\sim P_B$, and where $P_B$ denotes the distribution of the type class of
$\cC_B$ which contains the most elements. This type class is denoted by $\cC'_B$
in the sequel.

An important observation used to prove \eqref{ratei} and \eqref{divi}
is that because $\R$ can be assumed to be strictly positive (or there
is nothing to prove), the set of non-$\star$ symbols of each codeword in
$\cC_B$
has at most $O(B)$ elements.

(Note that $P_B$ may vary as a function of $\nu$. However, for ease
of exposition, we assume that $P_B$ is the same for all $\nu$. This assumption
is without loss of generality, because we can group the $\nu$'s together based
on their associated $P_B$, and as will become apparent from the analysis,
our arguments can be applied to each group separately. Since $A=2^{\beta B}$, for subsets
containing at least $A 2^{-\sqrt{B}}$ $\nu$'s, our arguments will be valid since
$\liminf_{B\to \infty} (1/B)\log(A 2^{-\sqrt{B}})=\beta $.
For $P_B$'s associated with fewer than this many $\nu$'s, since there are only
a polynomial number of $P_B$'s, the probability of $\nu$ having any such
$P_B$ is $o(1)$.)

\subsection{Proof of \eqref{ratei} and \eqref{ratei2}} The intuition for these inequalities is
that an asynchronous code must also be good for the synchronous channel,
and hence a suitable notion of rate is bounded by the
synchronous channel capacity. Formally, ${\cal{C}}_B'$ is clearly a good code
for the synchronous
channel, {\it{i.e.}}, if we reveal $\nu$ to the receiver
and decoding happens at time $\nu+\cD_B$, it is
possible to achieve an error probability bounded away from $1$ whenever $B$ is large enough.
From the strong converse for synchronous communication (see, {\it{e.g.}}, \cite[Corollary 6.4, p. 87]{CK}) it
follows that when $\star \in \cal{X}$ and $\star$ has zero cost, for any $\eta>0$,
\begin{align}\label{dBa}
\frac{\log |{\cal{C}}_B'|}{\cD_B}\leq I(X;Y)(1+\eta/2)
\end{align}
for all $B$ large enough. Similarly, when $\star \not \in \cal{X}$,
for any $\eta>0$,
\begin{align}\label{dB}
\frac{\log |{\cal{C}}_B'|}{n}\leq I(X;Y)(1+\eta/2) + \frac{\delta B}{n}
\end{align}
for all $B$ large enough, where $n$ denotes the number of non-$\star$ symbols in each
codeword. This can be seen by observing that the codewords
can be classified according to the value of $\sigma$, and for a given 
$\sigma$, only a rate of $I(X;Y)(1+\eta/2)$ can be supported. Because
of the delay constraint, only $2^{\delta B}$ choices of $\sigma$ are possible.

Now, since the number of
non-$\star$ symbols in any codeword is $O(B)$,
the number of possible types $P_B$ grows no faster than
polynomially with $B$.
To see this, note that there are $|\cX|$ input symbols, and we have $O(B)$
choices for the probability assigned to each
non-$\star$ symbol. Since there
is at most
one zero cost symbol (namely, the $\star$ symbol), $P_B$ is completely
determined by the number of occurrences of the
non-$\star$ symbols.
Thus, there are only a total of $O(B^{|\cX|})$ possible types $P_B$ satisfying
the constraint
of having $O(B)$ non-$\star$ symbols.
This implies that
$$\frac{\log
|{\cal{C}}_B'|}{\cD_B}=\frac{\log|{\cal{C}}_B|}{\cD_B}(1-o(1))\,$$
when $\star \in \cal{X}$, and similarly for the case when $\star \not \in \cal{X}$.
Combining this with \eqref{dBa} and \eqref{dB}, we obtain
\begin{align*}
\frac{\log |{\cal{C}}_B|}{\cD_B}\leq I(X;Y)(1+\eta)\,
\end{align*}
when $\star \in \cal{X}$, and 
\begin{align*}
\frac{\log |{\cal{C}}_B|}{n}\leq I(X;Y)(1+\eta)+\delta{B}/n\,
\end{align*}
when $\star \not \in \cal{X}$.
Note that $\log |{\cal{C}}_B|=B$ by definition.
Thus, by multiplying and dividing the left-hand sides of the above
inequalities by $K({\cal{C}}_B)$, and by noting that $K({\cal{C}}_B')\leq
K({\cal{C}}_B)$ by the definition of the cost of a code (see
Definition~\ref{costc} and recall that by definition, the extended codeword for
message $m$ has the same cost as $c^n(m)$), the above inequalities become 
\begin{align*}
\frac{K({\cal{C}}_B')}{\cD_B}\R\leq I(X;Y)(1+\eta)\,
\end{align*}
and
\begin{align*}
\frac{K({\cal{C}}_B')}{n}\R\leq \frac{I(X;Y)}{1-\delta}(1+\eta)\,.
\end{align*}
Since $K({\cal{C}}_B')=\cD_B \ex [k(X)]$ when $\star \in \cal{X}$
and $K({\cal{C}}_B')=n \ex [k(X)]$ when $\star \not\in \cal{X}$, inequalities
\eqref{ratei} and \eqref{ratei2} follow. 
Hence, if  \eqref{ratei} or \eqref{ratei2}, as appropriate, doesn't hold, then
the maximal error probability tends to one.

\subsection{Proof of \eqref{divi}}

We show that if  inequality \eqref{divi} is reversed, then a decoder that satisfies the delay constraint has an average error over messages that tends to one.  To prove this, we introduce the concepts
of ``effective output process'' and ``augmented decoder.''

\noindent{\emph{Effective output process:}} The
``effective'' output process is the random output process
``viewed'' by the sequential
decoder, {\it{i.e.}}, it is generated as if there were pure
noise after the transmission of the
{\emph{extended}} codeword. Specifically,  the distribution of the effective output process is as
follows. The $Y_i$'s for\footnote{Notice that because of
\eqref{ratei}, $d_B$ is a strictly positive
quantity.}
\begin{equation*}
i\in \left\{1,\ldots,\nu-1\right\}
\cup\left\{\nu+ d_B,\ldots,A_n+n-1\right\}
\end{equation*}
are i.i.d.\ according to $Q_\star$, whereas the block
\begin{equation*}
Y_\nu,Y_{\nu+1},\ldots,Y_{ \nu+d_B-1}
\end{equation*}
is distributed according to
$$\frac{1}{|\cC'_B|}\sum_{m}
Q(\cdot|c(m))\,,$$ the output
distribution given that a \emph{randomly
selected} (extended) codeword from $\cC'_B$ has been
transmitted.  With a slight abuse of notation, in the remainder
of the proof we use $Y_1,Y_2,\ldots,Y_{\asynclev+n-1}$ to denote the
effective output process.

\noindent{\emph{Augmented decoder:}}
An augmented decoder is a decoder which is revealed
the complete effective output sequence and, in
addition, is informed that the
message was sent
in one of \begin{align}\label{rr}
r_B\triangleq \left\lfloor \frac{A+n-1-\nu \mbox{
mod } \cD_B}{\cD_B}\right\rfloor
\end{align}
 consecutive (disjoint) blocks of
duration $\cD_B$, as shown in Fig.~\ref{graphees2}. Note that\footnote{We
use the notation $f(B)\doteq g(B)$ whenever the
functions $f$ and $g$ are exponentially equal, {\it{i.e.}}, if
$$\lim_{B\to \infty}\frac{1}{B}\log f(B)=\lim_{B\to
\infty}\frac{1}{B}\log g(B)\,.$$}
 \begin{align}\label{rrr}
r_B\doteq 2^{B(\beta-\delta)}\,.\end{align}
\begin{figure}
\begin{center}
\input{line}
\caption{\label{graphees2} Parsing of the entire
received sequence of size $A+n-1$
into $r_B$ blocks of
length $\cD_B$, one of which is
generated by the sent message, while the others are generated by pure
noise.}
\end{center}
\end{figure}
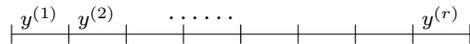
An augmented decoder, in addition to
outputting a message, also outputs an estimate
of the block of size $d_B$
corresponding to the time interval during which the message was sent.

Suppose the decoder of $\cC_B'$
achieves  (maximum)
communication delay less than $\cD_B$ with
probability equal to $1-\tilde{\varepsilon}_B$. Further, suppose it can output the correct message with
maximum error probability ${\varepsilon}_B$. 
Hence,  the corresponding augmented decoder can
both output  
the block of size $\cD_B$ which corresponds to
the actual transmission period, and output the
correct message, with maximum error probability
at most 
$\varepsilon_B+\tilde{\varepsilon}_B $. 
We now show that if \eqref{divi} doesn't hold,
then with
probability approaching one, pure noise will produce many output
blocks that look as if they were generated by
some codeword. This implies that
$\varepsilon_B+\tilde{\varepsilon}_B \to 1$.
Therefore, if the delay constraint is
satisfied with  $\tilde{\varepsilon}_B\to 0$,
then ${\varepsilon}_B\to 1$. Hence,  if the decoder of $\cC_B'$
achieves  (maximum)
communication delay less than $\cD_B$ with
probability tending to one, its error
probability will tend to one whenever 
\eqref{divi} doesn't hold.

To develop some intuition for proving \eqref{divi},
we first consider the simpler setting where
there is only a single message. We then
generalize to the multiple message
case to obtain \eqref{divi}.
 \subsubsection{Single message} Suppose there is only
one
codeword to be transmitted. The augmented decoder's only task is thus
to output the block of size $\cD_B$ that
corresponds to the period when $c(m)$ was
sent.


For this specific setting, we show
that if $\beta$
is sufficiently large, the decoder will not be able
to perform the task reliably, because the
noise is likely to produce several blocks that look as though
they were generated by $c(m)$.
More precisely, we show that the augmented decoder
has a large probability of error
(asymptotically equal to one) whenever for some
$\eta>0$ and all $B$ large enough,
\begin{align}\label{hcon}
B(\beta-\delta-\eta) > \cD_B D(XY||XY_\star)\,.
\end{align}

Let $\bar{c}(m)$ denote the extended codeword $c(m)$ without zero-cost symbols
and let $\bar{Y}(m)$ be its corresponding output. For instance, if the extended
codeword is $c(m)=1, 2, \star,2,\star$ and its
corresponding random output vector $Y(m)$ takes value $2,
2, 1,\star, 1$ then $\bar{c}(m)=1,2,2$ and $\bar{Y}(m)=2,2,\star$. Further,
let $\hat{Q}$ be the empirical distribution of $\bar{Y}(m)$ conditioned on 
 $\bar{c}(m)$, {\it{i.e.}}, $\hat{Q}$ satisfies
$$\hat{P}_{\bar{c}(m),\bar{Y}(m)}(x,y)={\bar{P}}_B(x)\hat{Q}(y|x),$$ where $\bar{P}_B$
denotes the empirical distribution of $\bar{c}(m)$.

The above restriction to the non-$\star$
symbols allows us to treat the various possible
delays---linear in $B$, subexponential in $B$, and exponential in $B$---in a
unified way.  Had we been interested only in the linear case, the argument would
also hold without the restriction to non-$\star$ symbols.

For a given fixed conditional probability distribution $\tilde{Q}$, denote by
$Z(m,\tilde{Q})$ the binomial random variable which represents the number of
pure noise blocks, out of $r_B-1$ of them, whose conditional empirical
distribution with respect to the non-$\star$ symbols of $\bar{c}(m)$ is
$\tilde{Q}$. Then the error probability of the augmented decoder can be lower
bounded as

\begin{align}\label{erlbo}
&\pr_m(\E)\nonumber \\
&\geq \sum_{\{\tilde{Q}:\tilde{Q}\approx
Q\}}\pr_m(\E|\hat{Q}=\tilde{Q})\times\pr_m(\bar{Y}(m)\in \cT_{\tilde{Q}}(\bar{c}(m))),
\end{align}
where the $\tilde{Q}$'s in the summation are conditional distributions
that are close to the actual channel $Q$.
Specifically,  $\tilde{Q}(\cdot|x)$ is such that
\begin{align}\label{qtilde}
||\tilde{Q}(\cdot|x)-Q(\cdot|x)||\leq 1/\log B
\end{align}
for any symbol $x\ne \star$ that appears
appears in $\bar{c}$
at least $\sqrt{B}$ times. And for any $x$ that
appears in $\bar{c}(m)$
less than $\sqrt{B}$ times,
$\tilde{Q}(\cdot|x)$ is arbitrary.

Now, conditioned on
$\{\hat{Q}=\tilde{Q}\}$, there are
$Z(m,\tilde{Q})$ pure noise blocks which look
statistically identical to the block
corresponding to the sent codeword, because
the empirical conditional distribution
of (the non-$\star$ codeword symbol positions of) each block is a sufficient statistic for
estimating the position of the sent codeword.
Hence, the augmented decoder fails
with probability at least
$$\ex\left(\frac{Z(m,\tilde{Q})}{Z(m,\tilde{Q})+1}\right)\,.$$
Therefore, from \eqref{erlbo},
 \begin{align}\label{erlboo}
&\pr_m(\E)\nonumber \\
&\geq \sum_{\{\tilde{Q}:\tilde{Q}\approx
Q\}}\ex\left(\frac{Z(m,\tilde{Q})}{Z(m,\tilde{Q})+1}\right)\pr_m(\bar{Y}(m)\in
\cT_{\tilde{Q}}(\bar{c}(m)))\,.
\end{align}

From Fact~\ref{fact:2},  the probability that
one single pure noise  block induces the
joint type $\bar{P}_B\tilde{Q}$ with $\bar{c}(m)$ is
\begin{align}\label{doteg}
\doteq
2^{-\bar{d}_BD(\bar{X}\tilde{Y}||\bar{X}Y_\star)}\doteq2^{-{d}_BD({X}{Y}||{X}Y_\star)}
\end{align}
where $\bar{X}\sim  \bar{P}_B$, where
$\bar{d}_B$ denotes the number of non-$\star$
symbols in $c(m)$. Note that the second
equality in \eqref{doteg} holds uniformly over
the set  $\{\tilde{Q}:\tilde{Q}\approx
Q\}$ by the continuity of divergence.\footnote{See foonote \ref{footnote:cont}.}

Therefore,
\begin{align}\label{doeeq}\ex(Z(m,\tilde{Q}))
&\doteq\frac{A}{\cD_B}2^{-\cD_BD(XY||XY_\star)}.
\end{align}
Since
$A=2^{\beta B}$, from \eqref{delcs}, \eqref{hcon}, and \eqref{doeeq} we get
$$\ex_m(Z(m,\tilde{Q}))\doteq 2^{\eta B}.$$ Since
$Z(m,\tilde{Q})$ is a binomial random variable, it can easily be seen
from Chebyshev's inequality (or the Chernoff bound) that
$Z(m,\tilde{Q})$ must be concentrated near its mean, from which
it follows that
\begin{align}\label{eriboo}
\ex_m\left(\frac{Z(m,\tilde{Q})}{Z(m,\tilde{Q})+1}\right)=1-o(1)\quad B\to
\infty\,.
\end{align}
From \eqref{erlboo} and \eqref{eriboo} we get
\begin{align}\label{erlbooo}
\pr_m(\E)&\geq (1-o(1))\sum_{\{\tilde{Q}:\tilde{Q}\approx
Q\}}\pr_m(\bar{Y}_\nu^{\nu+\bar{d}_B-1}\in
\cT_{\tilde{Q}}(\bar{c}(m)))\nonumber \\
&= 1-o(1)
\end{align}
as $B\to \infty$,
where the second equality follows from Chebyshev's inequality.
We conclude that for the single message case, the
error probability tends to one whenever
\eqref{hcon} holds.

\subsubsection{Multiple messages}
The main additional ingredient used to establish~\eqref{divi} is the fact that
the decoder does not know
a priori the transmitted message. Because of
this, the augmented decoder's task is
more difficult to perform; pure noise can induce an
error whenever it generates a block that is
typical with {\emph{any}} of the (extended) codewords from
$\cC_B'$. The key
element in the analysis consists in showing that the
``typicality'' regions associated with different
codewords are essentially disjoint, {\it{i.e.}}, that
the probability of the noise generating a block typical with any
message is essentially $|\cC_B'|$ times the probability for
the single message case. This,
together with the above argument for the single message
case, yields the desired result.

Observe that since ${\cal{C}}_B'$ achieves a maximum error probability
on the asynchronous channel that is less than
$\varepsilon_B$, the (extended) codewords
${\cal{C}}_B'$ can also achieve a  maximum error probability
on the synchronous channel that is less than
$\varepsilon_B$---if we reveal $\nu$ to the
decoder, the channel becomes synchronous, and the error probability does not
increase. Therefore, assuming that the decoder is
deterministic, we can assign \emph{disjoint} decoding
regions $D(m)$ to each  codeword
of ${\cal{C}}_B'$ such that, with probability at least $1-\varepsilon_B$, after
transmission
over the synchronous channel $Q$, the channel output lies in the decoding region
$D(m)$ assigned to the transmitted codeword $c(m)$.
If the decoder of ${\cal{C}}_B'$ is randomized,
one can easily construct an expurgated code with a deterministic decoder and asymptotically the same rate as follows. Since the maximum error probability of ${\cal{C}}_B'$  is at most $\varepsilon_B$, the average error probability is at most $\varepsilon_B$, hence the average error probability under MAP decoding is also at most $\varepsilon_B$ (note that MAP decoding minimizes the average error probability, not necessarily the maximum error probability). Now, without loss of optimality, the MAP decoder can be restricted to be deterministic. If we remove the half of the codewords with the largest error probability, we remain with a code whose maximum error probability is at most $2\varepsilon_B$ under a deterministic (MAP) decoding. This expurgated code and its decoding regions $\{D(m)\}$ can now be used for the argument.

Adapting the argument used for the single
message case, fix a conditional distribution
$\tilde{Q}\approx Q$ (see \eqref{qtilde}), and
let $Z(m,\tilde{Q})$ denote the binomial random
variable representing the number of pure noise
blocks that induce the conditional empirical
distribution $\tilde{Q}$ with $\bar{c}(m)$. For each message
$m$, define $D(m,\tilde{Q})$ as the
intersection of the decoding region $D(m)$
with  $\cT_{\tilde{Q}}(\bar{c}(m))$---that is
the set of sequences $y_1,y_2,\ldots, y_{d_B}$
in $D(m)$ whose $y_i$'s corresponding to the
non-$\star$ symbols of $c(m)$ have an empirical
distribution $\tilde{Q}$ given $\bar{c}(m)$. Note that since
the decoding regions are disjoint, the sets
$D(m,\tilde{Q})$ are also disjoint.

Define
$$Z(\tilde{Q})\triangleq \sum_{m}Z(m,\tilde{Q}),$$
and
$$D(\tilde{Q})\triangleq \cup_{m}D(m,\tilde{Q})\,.$$
Then,
\begin{align}\label{zq}
\ex&[Z(\tilde{Q})] =\sum_m\ex[Z(m,\tilde{Q})]\nonumber \\
& =  (r_B-1) \sum_{m} \pr_{\star}(D(m,\tilde{Q}))\nonumber \\
& = (r_B-1)\sum_{m} 2^{-{\cD}_B
(D({X}{Y}||{X}Y_\star)+o(1))}\pr_{m}(D(\tilde{Q},m))\nonumber\\
& =  \frac{A}{\cD_B} 2^{-\cD_B D(X{Y}||XY_\star)(1+o(1))}\sum_{m}
\pr_m(D(m,\tilde{Q}))\nonumber \\
& =  \frac{A}{\cD_B} 2^{-\cD_B D(X{Y}||XY_\star)(1+o(1))}2^B
\pr(D(M,\tilde{Q})),
\end{align}
where $\pr_{\star}$ denotes  the
output distribution corresponding to 
$\bar{d}_B$ symbols $\star$; and where
$\pr_{m}$
denotes the output distribution when the channel input is
 $\bar{c}(m)$.

The first equality in \eqref{zq} follows from the definition of $Z(\tilde{Q})$.
The second equality follows from the definition of $Z(m,\tilde{Q})$ and the
fact that there are $r_B-1$ pure
noise blocks (see \eqref{rr}). The third equality in \eqref{zq} holds since the
probability under $\pr_{\star}$ of any sequence in
$D(\tilde{Q},m)$
is equal to $2^{-{\cD}_B
(D({X}Y||{X}Y_\star)+o(1))}$ times the probability of that
sequence under $\pr_{m}$. To see this
note that for any $y\in D(\tilde{Q},m)$ we have \cite[Lemma 2.6]{CK}
$$\pr_{m}(y)=2^{-\bar{d}_B(H(\tilde{Y}|\bar{X})+D(\bar{X}\tilde{Y}||\bar{X}Y))}$$
and
$$\pr_{\star}(y)=2^{-\bar{d}_B(H(\tilde{Y})+D(\tilde{Y}||Y_\star))}\,.$$
Hence,
\begin{align*}\pr_{\star}(y)&=\pr_{m}(y)2^{-\bar{d}_B(D(\bar{X}\tilde{Y}||\bar{X}Y_\star)-D(\bar{X}\tilde{Y}||\bar{X}Y))}\nonumber
\\
&=\pr_{m}(y)2^{-{d}_B(D(XY||XY_\star)+o(1))},\nonumber
\end{align*}
since $\tilde{Q}\approx Q$, by continuity of divergence.\footnote{See footnote \ref{footnote:cont}.}
The fourth equality  in \eqref{zq} follows from
\eqref{rrr}. For the  fifth inequality in \eqref{zq} we defined
$$\pr( D(M,\tilde{Q}))\triangleq \frac{1}{|\cC'_B|} \sum_{m}
\pr_{m}(D(m,\tilde{Q}))\,,$$ the average  probability of successful decoding of
the code $\cC'_B$ and having an input/output joint type equal to
$P_B\tilde{Q}$---in the above definition, $M$ denotes the random message to be
transmitted.

Now, recall that the probability of successful decoding of $\cC'_B$ is at least
$1-\varepsilon_B$ (see paragraph after \eqref{rr}), hence
$$\ex_{\hat{Q}}\pr( D(M,\hat{Q}))\geq 1-\varepsilon_B\,.$$
Therefore, by Markov's inequality,
\begin{align}
\pr(\hat{Q}: \pr\big(D(M,\hat{Q}))\geq 1-\sqrt{\varepsilon_B}\big)&\geq
1-\sqrt{\varepsilon_B}\,,\nonumber
\end{align}
{\it{i.e.}}, with probability $1-\sqrt{\varepsilon_B}=1-o(1)$, the empirical
channel $\hat{Q}$ yields a probability of successful decoding
$1-\sqrt{\varepsilon_B}=1-o(1)$.
Denoting by $\{\tilde{Q} \sim Q\}$ the set of conditional distributions
$\tilde{Q}$ such that $\tilde{Q}\approx Q$ (see \eqref{qtilde}) and such that
$$\pr (D(M,\hat{Q}))\geq 1-\sqrt{\varepsilon_B}\,,$$
it follows that
\begin{align}\label{prq1}
\pr(\hat{Q} \sim Q)=1-o(1),
\end{align}
since $\pr(\hat{Q} \approx Q)=1-o(1)$. Hence, from \eqref{zq} and \eqref{rrr}, we
get \begin{align}
\label{zfq}
\ex&[Z(\tilde{Q})] =  2^{B(\beta-\delta)} 2^{-\cD_B
D(X{Y}||XY_\star)(1+o(1))}2^B
\end{align}
uniformly over $\{\tilde{Q}\sim Q\}$. Hence, if for some $\eta>0$ we have
\begin{align}\label{iol}
B(1+\beta-\delta-\eta) >\cD_BD(XY||XY_\star)\,,
\end{align}
 then $\ex
Z(\tilde{Q})\doteq 2^{\eta B }$, and using that
$Z(\tilde{Q})$ is a binomial random variable, we get
$$\ex\left(\frac{Z(\tilde{Q})}{Z(\tilde{Q})+1}\right)=1-o(1)\,.$$

Proceeding as in \eqref{erlboo}, the error
probability (averaged over messages) of the augmented decoder  is
lower bounded as
\begin{align}\label{erlbouf}
&\bar{\pr}(\E)\geq \sum_{\{\tilde{Q}:\tilde{Q}\sim
Q\}}\bar{\pr}(\E|\hat{Q}=\tilde{Q})\pr(\bar{Y}(M)\in
\cT_{\tilde{Q}}(\bar{c}(M)))\nonumber
\\
&\geq \sum_{\{\tilde{Q}:\tilde{Q}\sim
Q\}}\ex\left(\frac{Z(\tilde{Q})}{Z(\tilde{Q})+1}\right)\pr(\bar{Y}(M)\in
\cT_{\tilde{Q}}(\bar{c}(M)))
\nonumber \\
&=(1-o(1))\,.
 \end{align}
Hence, if  \eqref{iol} holds for some $\eta>0$, or, equivalently, if
$$\bm{R}\ex (k(X))(1+\beta-\delta-\eta) >D(XY||XY_\star)$$
since  $B/d_B=\bm{R}\ex (k(X))$, the error probability tends to one as $B\to
\infty$. This implies that if a code achieves rate $\bm{R}>0$ at timing
uncertainty per information bit $\beta$ and delay exponent $\delta$ then \eqref{divi} holds. This completes the proof of the
converses for Theorems~\ref{ulimit} and \ref{ulimit3}.
\end{IEEEproof}
\begin{IEEEproof}[Converse of Theorem~\ref{ulimit4}]
The converse proof for Theorem~\ref{ulimit4} is almost the same as the
the converse proof for Theorem~\ref{ulimit}. As for the achievability
proofs, the main idea is to find a suitable replacement for the set
$\{1,2,\ldots,A\}$
of time slots that the receiver needs to consider. For the proof, we choose the
set
of time slots as a function of the coding scheme under consideration.
In more detail, given any reliable coding scheme, {\it{i.e.}}, any coding scheme
for
which the probability of error $\varepsilon_B \rightarrow 0$ as $B \rightarrow
\infty$,
for each value $t$, consider the probability that the decoder
makes an error or has delay greater than $\cD_B$ conditioned on the event
$\nu=t$.
We will replace the set $\{1,\ldots, A\}$ with the set
${\cal{S}}(\sqrt{\varepsilon_B})$
of times $t$ for which this conditional
probability is at most $\sqrt{\varepsilon_B}$.
Observe that the conditional probability of error, averaged over $\nu$, is by
definition at most $2\varepsilon_B$,
so Markov's inequality says that the probability (over the distribution of
$\nu$)
that this conditional probability is larger than $\sqrt{\varepsilon_B}$ is at
most
$2\sqrt{\varepsilon_B}$. Thus, $\nu$ is in ${\cal{S}}(\sqrt{\varepsilon_B})$
with
probability at least $1-2\sqrt{\varepsilon_B}$.
The key property of this construction is that the decoder
for the given coding scheme can with high probability correctly
decode the message within a delay
of $\cD_B$ for \emph{each member} of
${\cal{S}}(\sqrt{\varepsilon_B})$.

We now apply the converse proof of Theorem~\ref{ulimit} to the set
${\cal{S}}(\sqrt{\varepsilon_B})$. First, we need to parse the output sequence
appropriately,
{\it{i.e.}}, split the output sequence into disjoint blocks of length $\cD_B$.
Recall that $r_B$,
the number of such disjoint blocks, was roughly $\frac{A}{\cD_B}$ in the
converse
proof of Theorem~\ref{ulimit}. Now, however, since
${\cal{S}}(\sqrt{\varepsilon_B})$ can be arbitrary, it is possible that
${\cal{S}}(\sqrt{\varepsilon_B})$ does not even contain any time slots congruent
to, say, $0$ mod $\cD_B$. To get around this minor technicality, observe that
by the pigeonhole principle, for at least
one value $x$ mod $\cD_B$, ${\cal{S}}(\sqrt{\varepsilon_B})$ contains at least
$\frac{|{\cal{S}}(\sqrt{\varepsilon_B})|}{\cD_B}$ time slots congruent to $x$
mod $\cD_B$.
For such an $x$, we choose $\nu$ uniformly from those elements in
${\cal{S}}(\sqrt{\varepsilon_B})$ that are congruent to $x$ mod $\cD_B$.
Because the decoder for the given coding scheme can with high probability
correctly
decode the message within a delay of $\cD_B$ for each member of
${\cal{S}}(\sqrt{\varepsilon_B})$, it follows that this decoder
can decode the message and determine the value of $\nu$ with high probability
even when $\nu$ is chosen as above.

From this point, we follow the converse proof of Theorem~\ref{ulimit}, with $r_B$
replaced
by $\frac{|{\cal{S}}(\sqrt{\varepsilon_B})|}{\cD_B}$ (equivalently, $A$ is
replaced
by the size of ${\cal{S}}(\sqrt{\varepsilon_B})$). At the end, we see that
a reliable decoder can exist only if for any $\eta>0$ and $B$ large enough,
$$B\left(1+\frac{\log({\cal{S}}(\sqrt{\varepsilon_B}))}{B}\right) \leq
\cD_B(D(XY|| XY_{\star})+\eta).$$
Thus, $\frac{\log({\cal{S}}(\sqrt{\varepsilon_B}))}{B}$ has replaced the role
played
by $\beta$ in the converse proof of Theorem~\ref{ulimit}. Finally, since
$\varepsilon_B \rightarrow 0$, $\sqrt{\varepsilon_B} \rightarrow 0$, so by
definition
of $\bar{\beta}$ $$\bar{\beta} \leq \limsup
\frac{\log({\cal{S}}(\sqrt{\varepsilon_B}))}{B},$$
completing the proof.
\end{IEEEproof}
\begin{IEEEproof}[Proof of Theorem \ref{ulimit2}]
Starting from Theorem \ref{ulimit},
\begin{equation}
\label{eq:distr} \bm{C}(\beta) = \max_X \min \left \{
\frac{I(X;Y)}{\ex[\ck(X)]}, \frac{I(X;Y) +
D(Y||Y_\star)}{\ex[\ck(X)](1+\beta)} \right \}.
\end{equation}

A simple upper bound is

\begin{eqnarray}
\bm{C}(\beta) & \le &  \max_X \frac{I(X;Y) +
D(Y||Y_\star)}{\ex[\ck(X)](1+\beta)} \\
& = & \frac{1}{1+\beta} \max_X \frac{\ex[f(X)]}{\ex[\ck(X)]},
\label{eq:ubc}
\end{eqnarray}
where $f(x)$ is the divergence between the distribution of $Y$
conditioned on $X=x$ and the distribution of $Y$ conditioned on
$X=\star$.

Using the fact that for nonnegative $a, b, c,$ and $d$ (with a suitable
convention for the case where $c$ and/or $d$ is $0$)
$$ \frac{a+b}{c+d} \le \max\left(\frac{a}{c}, \frac{b}{d}\right),$$
we see that the above maximum is achieved for an input
distribution with a point mass at $a^*$, where

$$ a^* = {\rm argmax}_x \frac{f(x)}{\ck(x)}.$$

However, the maximizing solution is not unique. Since $f(\star) =
\ck(\star) = 0$,
$$ \frac{p f(\star) + (1-p)f(a^*)}{p \ck(\star) + (1-p) \ck(a^*)} =
\frac{f(x)}{\ck(x)}$$ for any $p \in [0,1]$. Hence, any input
distribution with two point masses, one at $\star$ and one at $a^*$,
will do. Going back to (\ref{eq:ubc}), we get

$$ \bm{C}(\beta) \le \frac{1}{1+\beta} \max_x \frac{f(x)}{\ck(x)}.$$

This upper bound is obtained by choosing the input distribution to
maximize the second term in the minimum of (\ref{eq:distr}). To
prove that this upper bound can be achieved, choose $X$ to have a
distribution with probability $p$ of being $\star$, and probability
$1-p$ of being $a^*$, where $p \rightarrow 1$. The first term in the min
approaches
$$ \max_x \frac{f(x)}{\ck(x)}$$
by Theorem 3 of \cite{Ve2}. The second term is
$$ \frac{1}{1+\beta} \max_x \frac{f(x)}{\ck(x)},$$
as derived above (true actually for any $p$, not only $p \rightarrow
1$). So, the second term is smaller, and we are always limited by the
timing uncertainty. This proves the desired result.
\end{IEEEproof}

\begin{rem}
Our results hold under the assumption that the only possible zero cost symbol is
the $\star$ symbol. The other cases, which we now briefly discuss, can be
handled with arguments similar to the ones used in this paper. 
\begin{itemize}
\item Two symbols in $\cX$ have zero cost: the capacity per unit cost is readily
seen to be infinite.
\item $\star\in\cal{X}$ and all $x\in \cX$ have positive cost: the
analysis in this paper can be applied, but would require some
slightly cumbersome notation. 
\item There is a single zero cost symbol $x\in \cX$ different than~$\star$: in
this case the asynchronous capacity per unit cost is
$$\bm{C}(\beta,\delta)=\frac{\bm{C}(0,0)}{1-2\delta} \quad 0\leq \delta\leq
\beta/2, \: \beta\geq 0\,,$$
{\it{i.e.}}, it is the synchronous capacity per unit cost multiplied by a factor
$1/(1-2\delta)$. 

The first thing to note in the above capacity expression is that it does not depend
on $\beta$. The reason for this is that no matter how large $\beta$ is it is always
possible to append to each codeword a long enough zero cost preamble that
guarantees the decoder is able to identify $\sigma$ with high probability. 

For an intuitive justification for the $1-2\delta$ factor, observe that in the
achievability proof of Theorem~\ref{ulimit3} case $b.$, $\delta B$ bits are
encoded via $\sigma$, the {\emph{start}} information time. When a symbol
different than $\star$ has zero cost, not only it is possible to encode
information through the start information time, but also in the codeword
``length.'' By codeword length we mean the time between $\sigma$ and the time of
the last non-zero cost symbol of the sent codeword. This allows to communicate
$2\delta B$ of information only through timing.
\end{itemize}
\end{rem}

\begin{IEEEproof}[Proof of Theorem \ref{gauss}]
A simple quantization argument can be used to derive
Theorem~\ref{gauss} from Theorem~\ref{ulimit}. For
achievability, one quantizes the input and
the output
real values to a finite alphabet.
Then, the achievability part of Theorem~\ref{ulimit} 
can be applied to this
quantized channel. Finally, take the limit of infinitely fine quantization to
proves that the stated rate is achievable.

For the converse, one adapts the method of types
by quantizing the set of probability distributions,
{\it{i.e.}}, one defines a type as a set of probability distributions
that are ``close'' to each other. With such a notion of type,
the converse part of Theorem~\ref{ulimit} can be applied,
and in the limit of infinitely fine quantization, one obtains
the desired converse result.
\end{IEEEproof}

\bibliographystyle{amsplain}
\bibliography{../../../../common_files/bibiog}

\begin{IEEEbiographynophoto}{Venkat Chandar}
received S.B. degrees in EECS and mathematics in 2006, an M. Eng.
in EECS in 2006, and a Ph. D. in EECS in 2010, all from MIT. His current
research interests include coding theory and algorithms, with an
emphasis on the construction and analysis of sparse graph codes for various
problems related to communication, compression, sensing, and
information-theoretic secrecy.
\end{IEEEbiographynophoto}
\begin{IEEEbiographynophoto}{Aslan Tchamkerten}
received the Engineer Physicist Diploma in 2000 and
the Ph.D. degree in Communications in 2005, both from the Ecole
Polytechnique F\'ed\'erale de Lausanne (EPFL), Switzerland. Between 2005 and
2008, he was a Postdoctoral Associate in the Department of Electrical
Engineering and Computer Science, Massachusetts Institute of Technology
(MIT), Cambridge. In 2008 he joined Telecom ParisTech (ex. Ecole
Nationale Sup\'erieure des T\'el\'ecommunications, ENST), Paris, France, where he
is currently Associate Professor. In 2009, he won
a junior excellence chair grant from the French National Research Agency
(ANR). His research interests are in information theory, applied Statistics,
and algorithms.
\end{IEEEbiographynophoto}
\begin{IEEEbiographynophoto}{David Tse} received the B.A.Sc. degree in systems
design engineering from University of Waterloo in 1989, and the M.S. and Ph.D.
degrees in electrical engineering from Massachusetts Institute of Technology
in 1991 and 1994 respectively. From 1994 to 1995, he was a postdoctoral member
of technical staff at A.T. \& T. Bell Laboratories. Since 1995, he has been at
the Department of Electrical Engineering and Computer Sciences in the
University of California at Berkeley, where he is currently a Professor. He
received a 1967 NSERC graduate fellowship from the government of Canada in
1989, a NSF CAREER award in 1998, the Best Paper Awards at the Infocom 1998 and
Infocom 2001 conferences, the Erlang Prize in 2000 from the INFORMS Applied
Probability Society, the IEEE Communications and Information Theory Society
Joint Paper Award in 2001, the Information Theory Society Paper Award in 2003,
the 2009 Frederick Emmons Terman Award from the American Society for
Engineering Education, and a Gilbreth Lectureship from the National Academy of
Engineering in 2012. He has given plenary talks at international conferences
such as ICASSP in 2006, MobiCom in 2007, CISS in 2008, and ISIT in 2009. He was
the Technical Program co-chair of the International Symposium on Information
Theory in 2004, and was an Associate Editor of the IEEE Transactions on
Information Theory from 2001 to 2003. He is a coauthor, with Pramod Viswanath,
of the text "Fundamentals of Wireless Communication", which has been used in
over 60 institutions around the world. \end{IEEEbiographynophoto}
\end{document}

%% file: canaltemps.tex
\setlength{\unitlength}{0.8bp}%
\begin{picture}(0,0)
\includegraphics[scale=0.8]{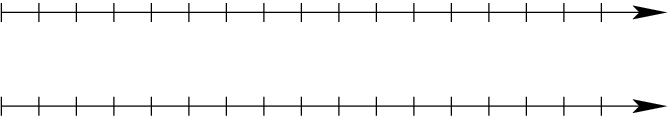}
\end{picture}%
\begin{picture}(321, 56)

\put(.7,-8){\scriptsize\makebox(0,0){$Y_1$}}
\put(18.7,-8){\scriptsize\makebox(0,0){$Y_2$}}
\put(54,-10){\makebox(0,0){$\ldots$}}
\put(.7,37){\makebox(0,0){$\star$}}
\put(18.7,37){\makebox(0,0){$\star$}}
\put(54,37){\makebox(0,0){$\ldots$}}
\put(90.2,37){\makebox(0,0){$\star$}}
\put(108.2,37){\tiny{\makebox(0,0){$c_1(m)$}}}
\put(108.2,65){\footnotesize{\makebox(0,0){$\nu$}}}

\put(153,37){\makebox(0,0){$\ldots$}}
\put(164,20){\footnotesize\makebox(0,0){$\tau_N$}}

\put(198,37){\tiny{\makebox(0,0){$c_N(m)$}}}
\put(216.5,37){\makebox(0,0){$\star$}}
\put(234.5,37){\makebox(0,0){$\star$}}
\put(261.5,37){\makebox(0,0){$\ldots$}}
\put(288.5,37){\makebox(0,0){$\star$}}

\end{picture}
\vspace{0.5cm}

%% file: line.tex
\setlength{\unitlength}{0.8bp}%
\begin{picture}(0,0)
\includegraphics[scale=0.8]{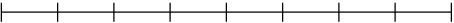}
\end{picture}%
\begin{picture}(218, 11)
\put(14,13){\makebox(0,0){\footnotesize{$y^{(1)}$}}}
\put(41,13){\makebox(0,0){\footnotesize{$y^{(2)}$}}}
\put(90,13){\makebox(0,0){{$\ldots\ldots$}}}

\put(203,13){\makebox(0,0){\footnotesize{$y^{(r)}$}}}

\end{picture}
\vspace{0.5cm}